\documentclass[final,5p,times,twocolumn]{elsarticle}

\usepackage{graphicx}
\usepackage{enumerate}
\usepackage{diagbox}
\usepackage{siunitx}

\usepackage{soul}
\usepackage[algo2e,linesnumbered,commentsnumbered, lined, boxed,ruled,vlined]{algorithm2e}
\def\HiLi{\leavevmode\rlap{\hbox to \hsize{\color{green!10}\leaders\hrule height .8\baselineskip depth .5ex\hfill}}}
\def\HiLiScalar{\leavevmode\rlap{\hbox to \hsize{\color{blue!10}\leaders\hrule height .8\baselineskip depth .5ex\hfill}}}

\usepackage[dvipsnames]{xcolor}

\definecolor{algoColorComment}{rgb}{0.5,0.5,0.5}

\SetCommentSty{mycommfont}

\usepackage{setspace}
\renewcommand{\imath}{\textrm{i}}

\SetAlFnt{\small}
\SetAlCapFnt{\small}
\SetAlCapNameFnt{\small}
\SetAlCapHSkip{0pt}
\SetKwProg{Fn}{function}{}{}
\SetKwBlock{Begin}{ }{ }

\usepackage{listings}
\definecolor{cppColorBackground}{rgb}{1.,1.,1.}
\definecolor{cppColorComment}{rgb}{0.0,0.67,1.}
\definecolor{cppColorLine}{rgb}{0.6,0.6,0.6}
\definecolor{cppColorString}{rgb}{0,0.501,145}
\definecolor{cppColorKey}{rgb}{0.5,0.5,0}
\definecolor{cppColorDigit}{rgb}{0,0,0.5}
\usepackage{pgfplots}
\pgfplotsset{compat=newest}
\usetikzlibrary{patterns}

\usepackage{subcaption}
\usepackage{hyperref}
\usepackage{amsmath}
\usepackage{amssymb}
\usepackage{makecell}
\usepackage{placeins} 

\newcommand{\codename}{{TurTLE}}%

\newcommand{\total}{{\mathrm{d}}}

\newcommand{\nameOfCAPList}{\ensuremath{\mathcal{L}}}%
\newcommand{\nameOfAction}{\ensuremath{\mathcal{A}}}%
\newcommand{\numberGridNodes}{\ensuremath{N}}%
\newcommand{\numberParticles}{\ensuremath{N_p}}%
\newcommand{\numberMPIProcesses}{\ensuremath{P}}%
\newcommand{\numberParticleOutputProcesses}{\ensuremath{P_O}}%
\newcommand{\numberSlicesPerProcess}{\ensuremath{P_s}}%
\newcommand{\numberParticlesPerProcess}{\ensuremath{P_p}}%
\newcommand{\numberParticlesPerSlice}{\ensuremath{S_p}}%
\newcommand{\numberInterpolationKernelNodes}{\ensuremath{I}}%

\usepackage{mathtools}
\DeclarePairedDelimiter\ceil{\lceil}{\rceil}

\newcommand{\bs}[1]{{\boldsymbol{#1}}}
\newcommand{\uu}{{\bs{u}}}

\newcommand{\FT}{{\mathcal{F}}}
\newcommand{\NL}{{\mathcal{N}}}
\newcommand{\XX}{{\bs{X}}}
\newcommand{\xx}{{\bs{x}}}
\newcommand{\cc}{{\bs{c}}}

\usepackage{array}

\journal{Computer Physics Communications}

\begin{document}

\begin{frontmatter}

\title{An Efficient Particle Tracking Algorithm for Large-Scale Parallel
    Pseudo-Spectral Simulations of Turbulence}
\author[a,b,d]{Cristian C. Lalescu}
\author[c,b,e]{B\'erenger Bramas}
\author[b]{Markus Rampp}
\author[a,f]{Michael Wilczek}

\cortext[d]{Corresponding author.\\\textit{E-mail address:} Cristian.Lalescu@mpcdf.mpg.de}
\address[a]{Max Planck Institute for Dynamics and Self-Organization, 37077 G\"ottingen, Germany}
\address[b]{Max Planck Computing and Data Facility, 85748 Garching, Germany}
\address[c]{Inria Nancy -- Grand-Est, CAMUS Team, 54600 Villers-l\`es-NancyFrance}
\address[e]{ICube, ICPS Team, 67400 Illkirch-Graffenstaden, France}
\address[f]{Theoretical Physics I, University of Bayreuth, 95440 Bayreuth, Germany}

\begin{abstract}
Particle tracking in large-scale numerical simulations of turbulent flows presents one of the major bottlenecks in parallel performance and scaling efficiency. Here, we describe a particle tracking algorithm for large-scale parallel pseudo-spectral simulations of turbulence which scales well up to billions of tracer particles on modern high-performance computing architectures. We summarize the standard parallel methods used to solve the fluid equations in our hybrid MPI/OpenMP implementation. As the main focus, we describe the implementation of the particle tracking algorithm and document its computational performance. To address the extensive inter-process communication required by particle tracking, we introduce a task-based approach to overlap point-to-point communications with computations, thereby enabling improved resource utilization. We characterize the computational cost as a function of the number of particles tracked and compare it with the flow field computation, showing that the cost of particle tracking is very small for typical applications.

\end{abstract}


\end{frontmatter}

\section{Introduction}

Understanding particle transport in turbulent flows is fundamental to the problem of turbulent mixing \cite{taylor1922plms, kolmogorov1991prsl_a, yeung1989jfm, yeung2002arfm, toschi2009arfm, homann2009pop} and relevant for a wide range of applications such as dispersion of particles in the environment \cite{stohl2002ae, stohl2006jgra, behrens2012erl, haszpra2013npg}, the growth of cloud droplets through collisions \cite{shaw2003arfm, Bodenschatz970, devenish2012qjrms, grabowski2013arfm, pumir2016arcmp}, and phytoplankton swimming in the ocean \cite{durham2013ncomms, breier2018pnas,pujara2018jfm}. Direct numerical simulations (DNS) of turbulence are nowadays an established tool for investigating such phenomena and have a long history in scientific computing \cite{orszag1972prl, yeung1988jcp, yokokawa2002supercomputing, ishihara2016prf, buaria2017cpc}. DNS have become a major application and technology driver in high-performance computing since the scale separation between the largest and the smallest scales increases drastically with the Reynolds number $R_\lambda$, which characterizes the degree of small-scale turbulence~\cite{pope2000book}. Dimensional estimates of the required computational resources scale at least as $R_\lambda^6$~\cite{pope2000book}. Recent literature \cite{yeung2018prf}, however, shows that, due the occurrence of extremely small-scale structures, resolution requirements increase even faster than simple dimensional arguments suggest. Until today DNS have reached values of
up to $R_\lambda \approx 2300$~\cite{ishihara2007jfm, yeung2015pnas, ishihara2016prf}, still smaller than the latest experiments, which have reached $R_\lambda > 5000$~\cite{kuechler2019jsp}, or natural settings such as cumulus clouds, which show Reynolds numbers on the order of $10^4$ \cite{warhaft2002pnas}.
Hence DNS of turbulence will continue to be computationally demanding for the foreseeable future.

Due to the large number of grid points, practical implementations of DNS typically employ one- or two-dimensional domain decompositions within a distributed memory parallel programming paradigm.
While the numerical solution of the field equations is typically achieved with well-established methods, the efficient implementation of particle tracking within such parallel approaches still poses major algorithmic challenges.
In particular, particle tracking requires an accurate interpolation of the flow fields on distributed domains, and particles traversing the domain need to be passed on from one sub-domain/process to another.
As the Reynolds number increases, the number of particles required to adequately sample the turbulent fields needs to grow with the increasing number of grid points since this is a measure of the degrees of freedom of the flow. In addition higher-order statistics might be needed to address specific research questions, and thus the number of particles required for converged statistics increases as well \cite{yeung2002arfm, biferale2004prl, eyink2011pre, eyink2013nature, biferale2014jfm, johnson2015pof, lalescu2018jfm, lalescu2018njp}.
Overall, this requires an approach which handles the parallel implementation in an efficient manner for arbitrarily accurate methods.
One option explored in the literature is the use of the high-level programming concept of \emph{coarrays}, in practice shifting responsibility for some of the communication operations to the compiler, see \cite{buaria2017cpc}.
The general solution that we describe makes use of MPI and OpenMP for explicit management of hardware resources.
The combination of MPI~\cite{Gropp1999} and OpenMP~\cite{omp45_spec} has become
a de facto standard in the development of large-scale applications
\cite{goerler2011jcp, mininni2011pc, pekurovsky2012siamjsc, p3dfft, clay2017cpc, chatterjee2018tarang}.
MPI~\cite{walker1996mpi} is used for communication between processes and OpenMP to manage multiple execution threads over multicore CPUs using shared memory.
Separately, large datasets must be processed with specific data-access patterns to make optimal use of modern hardware, as explained for example in \cite{homann2018cpc}.

To address the challenges outlined above, we have developed the numerical framework ``Turbulence Tools: Lagrangian and Eulerian'' (\codename{}), a flexible pseudo-spectral solver for fluid and turbulence problems implemented in C++ with a hybrid MPI/OpenMP approach~\cite{turtle_repository, turtle_documentation}. \codename{} allows for an efficient tracking of various types of particles. In particular, \codename{} showcases a parallel programming pattern for particle tracking that is easy to adapt and implement, and which allows for efficient use of available resources. Our event-driven approach is especially suited for the case where individual processes require data exchanges with several other processes while also being responsible for local work. For this, asynchronous inter-process communication and tasks are used, based on a combined MPI/OpenMP implementation.
As we will show in the following, \codename{} permits numerical particle tracking at relatively small costs, while retaining flexibility with respect to number of particles and numerical accuracy. We show that \codename{} scales well up to $O(10^4)$ computing cores, with the flow field solver approximately retaining the performance of the used Fourier transform libraries for DNS with $3 \times 2048^3$ and $3 \times 4096^3$ degrees of freedom. We also measure the relative cost of tracking up to $2.2 \times 10^{9}$ particles as approximately only 10\% of the total wall-time for the $4096^3$ case, demonstrating the efficiency of the new algorithm even for very demanding particle-based studies.

In the following, we introduce \codename{} and particularly focus on the efficient implementation of particle tracking. Section~\ref{sec:numerical} introduces the evolution equations for the fluid and particle models, as well as the corresponding numerical methods. Section~\ref{sec:implementation} provides an overview of our implementation, including a more detailed presentation of the parallel programming pattern used for particle tracking. Finally, Section~\ref{sec:performance} summarizes a performance evaluation using up to 512 computational nodes.

\section{Evolution equations and numerical method}
\label{sec:numerical}

\subsection{Fluid equations}

While \codename{} is developed as a general framework for a range of fluid equations, we focus on the Navier-Stokes equations as a prototypical example in the following. The incompressible Navier-Stokes equations take the form
\begin{equation}
    \begin{aligned}
        \partial_t \bs u + \bs u \cdot \nabla \bs u &=
        - \nabla p
        + \nu \Delta \bs u + \bs f, \\
        \nabla \cdot \bs u &= 0.
    \end{aligned}
    \label{eq:NS}
\end{equation}
Here, $\bs u(\bs x, t)$ denotes the three-dimensional velocity field, $p(\bs x, t)$ is the kinematic pressure, $\nu$ is the kinematic viscosity, and $\bs F(\bs x, t)$ denotes an external forcing that drives the flow. We consider periodic boundary conditions, which allows for the use of a Fourier pseudo-spectral scheme.
Within this scheme, a finite Fourier representation is used for the fields, and the non-linear term of the Navier-Stokes equations is computed in real space --- an approach pioneered by Orszag and Patterson \cite{orszag1972prl}.
For the concrete implementation in \codename{}, we use the vorticity formulation of the Navier-Stokes equation, which takes the form
    \begin{equation}
            \partial_t {\bs \omega} =
            \nabla \times (\bs u \times \bs \omega )
            + \nu \Delta {\bs \omega} + \bs F,
        \label{eq:NSVE_real}
    \end{equation}
where $\bs \omega(\bs x, t) = \nabla \times \bs u(\bs x, t)$ is the vorticity field and $\bs F(\bs x, t) = \nabla \times \bs f(\bs x, t)$ denotes the curl of the Navier-Stokes forcing.
The Fourier representation of this equation takes the form \cite{homann2006phd, wilczek2010phd}
    \begin{equation}
            \partial_t \hat{\bs \omega} (\bs k, t) =
            \imath \bs k \times \FT [\bs u(\bs x, t) \times \bs \omega (\bs x, t) ]
            - \nu k^2 \hat{\bs \omega} (\bs k, t) + \hat{\bs F}(\bs k, t),
        \label{eq:NSVE}
    \end{equation}
where $\FT$ is the direct Fourier transform operator. In Fourier space, the velocity can be conveniently computed from the vorticity using Biot-Savart's law,
\begin{equation}
    \hat{\bs u} (\bs k, t) = \frac{\imath \bs k \times \hat{\bs \omega} (\bs{k}, t)}{k^2}.
\end{equation}

Equation \eqref{eq:NSVE} is integrated with a
third-order Runge-Kutta method \cite{shu1988jcp}, which  is an explicit
Runge-Kutta method with the Butcher tableau
    \begin{equation}
    \begin{array}
        {c|ccc}
        0\\
        1 & 1\\
        1/2 & 1/4 & 1/4 \\
        \hline
        & 1/6 & 1/6 & 2/3
        \end{array}.
        \label{eq:Butcher_tableau}
    \end{equation}
In addition to the stability properties described in \cite{shu1988jcp}, this method has the advantage that it is memory-efficient, requiring only two additional field allocations, as can be seen from \cite{homann2006phd}
    \begin{equation}
    \begin{aligned}
        \hat{\bs w}_1(\bs k) &= ({\hat{\bs \omega}}(\bs k, t) + h \NL[\hat{\bs \omega}(\bs k, t)])e^{-\nu k^2 h}, \\
        \hat{\bs w}_2(\bs k) &= \tfrac{3}{4}\hat{\bs \omega}( \bs k, t)e^{-\nu k^2 h/2}
        +
        \tfrac{1}{4}(\hat{\bs w}_1(\bs k) + h \NL[\hat{\bs w}_1(\bs k)])e^{\nu k^2 h/2}, \\
        \hat{\bs \omega}(\bs k, t+h) &= \tfrac{1}{3}\hat{\bs \omega}(\bs k, t)e^{-\nu k^2 h} +
        \tfrac{2}{3}(\hat{\bs w}_2(\bs k) + h \NL[\hat{\bs w}_2(\bs k)])e^{-\nu k^2 h/2},
    \end{aligned}
        \label{eq:efficient_RK3}
    \end{equation}
where $h$ is the time step, limited in practice by the Courant–Friedrichs–Lewy (CFL) condition \cite{CFL1967}. The nonlinear term
\begin{equation}
        \NL[\hat{\bs w} (\bs k)] =
        \imath \bs k \times \FT \left[\FT^{-1}\left[\tfrac{\imath \bs k \times
        \hat{\bs w} (\bs{k})}{k^2}\right] \times \FT^{-1}\left[\hat{\bs w} (\bs
        k)\right]\right]
        \label{eq:nonlinear_term_notation}
    \end{equation}
is computed by switching between Fourier space and real space.
If the forcing term is nonlinear, it can be included in the right-hand side of \eqref{eq:nonlinear_term_notation}.
To treat the diffusion term, we use the standard integrating factor technique \cite{CHQZ1988} in \eqref{eq:efficient_RK3}.

Equation \eqref{eq:NSVE} contains the Fourier transform of a quadratic nonlinearity. Since numerical simulations are based on finite Fourier representations, the real-space product of the two fields will in general contain unresolved high-wavenumber harmonics, leading to aliasing effects \cite{CHQZ1988}.
In \codename{}, de-aliasing is achieved through the use of a smooth Fourier filter, an approach that has been shown in \cite{hou2007jcp} to lead to good convergence to the true solution of a PDE, even though it does not completely remove aliasing effects.

 The Fourier transforms in \codename{} are evaluated using the FFTW library \cite{FFTW2005}. Within the implementation of the pseudo-spectral scheme, the fields have two equivalent representations: an array of vectorial Fourier mode amplitudes, or an array of vectorial field values on the real-space grid. For the simple case of 3D periodic cubic domains of size $[0, 2\pi]^3$, the real-space grid is a rectangular grid of $\numberGridNodes\times\numberGridNodes\times \numberGridNodes$ points, equally spaced at distances of $\delta \equiv 2\pi / \numberGridNodes$.
Exploiting the Hermitian symmetry of real fields, the Fourier-space grid consists of $\numberGridNodes \times \numberGridNodes \times (\numberGridNodes/2+1)$ modes. Therefore, the field data consists of arrays of floating point numbers, logically shaped as the real-space grid or arrays of floating point number pairs (e.g. \texttt{fftw\_complex}) logically shaped as the Fourier-space grid. Extensions to non-cubic domains or non-isotropic grids are straightforward.

The direct numerical simulation algorithm then has two fundamental constructions: loops traversing the fields, with an associated cost of $O(\numberGridNodes^3)$ operations, and direct/inverse Fourier transforms, with a cost of $O(\numberGridNodes^3 \log \numberGridNodes)$ operations.

\subsection{Particle equations}
\label{sec:particle_equations}

A major feature of \codename{} is the capability to track different particle types, including Lagrangian tracer particles, ellipsoids, self-propelled particles and inertial particles. To illustrate the implementation, we focus on tracer particles in the following.

Lagrangian tracer particles are virtual markers of the flow field starting from the initial position $\bs x$. Their position $\XX$ evolves according to
    \begin{equation}
        \tfrac{\total}{\total t} \XX(\xx, t) = \uu(\XX(\xx, t), t),
        \hskip .02 in
        \XX(\xx, 0) = \xx.
        \label{eq:tracers}
    \end{equation}

The essential characteristic of such particle equations is that they require as input the values of various flow fields at arbitrary positions in space.

\codename{} combines multi-step Adams-Bashforth integration schemes (see, e.g., \S6.7 in \cite{atkinson1989book}) with spline interpolations \cite{lalescu2010jcp} in order to integrate the ODEs. Simple Lagrange interpolation schemes (see, e.g., \S3.1 in \cite{atkinson1989book}) are also implemented in \codename{} for testing purposes. There is ample literature on interpolation method accuracy, efficiency, and adequacy for particle tracking, e.g. \cite{yeung1988jcp, lekien2005ijnme, homann2007cpc, vanhinsberg2012siamjsp, vanhinsberg2013pre}.
The common feature of all interpolation schemes is that they can be represented as a weighted real-space-grid average of a field, with weights given by the particle's position.
For all practical interpolation schemes, the weights are zero outside of a relatively small kernel of grid points surrounding the particle, i.e.~the formulas are ``local''.
For some spline interpolations, a non-local expression is used, but it can be rewritten as a local expression where the values on the grid are precomputed through a distinct global operation \cite{yeung1988jcp} --- this approach, for example, is used in \cite{buaria2017cpc}.

Thus an interpolation algorithm can be summed up as follows:
    \begin{enumerate}
        \item compute $\widetilde{\XX} = \XX \mod 2\pi$ (because the domain is periodic).
        \item find the closest grid cell to the particle position $\widetilde{\XX}$, indexed by $\bs c \equiv (c_1, c_2, c_3)$.
        \item compute $\bar{\xx} = \widetilde{\XX} - \cc \delta$.
        \item compute a sum of the field over a kernel of
            $\numberInterpolationKernelNodes$
            grid points in each of the 3 directions, weighted by some
            polynomials:
            \begin{equation}
                \uu(\XX) \approx
                \sum_{i_1, i_2, i_3=1-\numberInterpolationKernelNodes/2}^{\numberInterpolationKernelNodes/2}
                \beta_{i_1}\left(\tfrac{\bar{x}_1}{\delta}\right)
                \beta_{i_2}\left(\tfrac{\bar{x}_2}{\delta}\right)
                \beta_{i_3}\left(\tfrac{\bar{x}_3}{\delta}\right)
                \uu(\bs c + \bs i) ,
                \label{eq:triple_sum}
            \end{equation}
    \end{enumerate}
    where $\bs i = (i_1,i_2,i_3)$.
    The cost of the sum itself grows as
    $\numberInterpolationKernelNodes^3$, the cube of the \emph{size of the interpolation kernel}.
    The polynomials $\beta_{i_j}$ are determined by the interpolation scheme (see \cite{lalescu2010jcp}).

    Accuracy improves with increasing $\numberInterpolationKernelNodes$. In \codename{}, interpolation is efficiently implemented even at large $\numberInterpolationKernelNodes$.
    As discussed below in \S\ref{sec:particletracersystem}, this is achieved by sorting particle data such that a minimal number of MPI messages is used to complete all the triple sums, independently of the total number of particles.

\section{Implementation}

\label{sec:implementation}
\subsection{Overview}

\codename{} follows the object-oriented paradigm to facilitate maintenance and extensions through class inheritance and virtualization.
The solver relies on two types of objects.
Firstly, all solvers inherit three basic elements from an abstract parent class: generic \emph{initialization}, \emph{do work} and \emph{finalization} functionality.
Solvers are then grouped in this first type of object, i.e.~members of the resulting class hierarchy.
The second type of object encapsulates essential data structures (i.e.~fields, sets of particles) and associated functionality (e.g. HDF5-based I/O): these are ``building block''-classes.
Each solver then consists of a specific ``arrangement'' of the building blocks.

The parallelization of \codename{} is based on a standard, MPI-based, one-dimensional domain-decomposition approach: The three-dimensional fields are decomposed along one of the dimensions into a number of slabs, with each MPI process holding one such slab. In order to efficiently perform the costly FFT operations with the help of a high-performance numerical library such as FFTW, process-local, two-dimensional FFTs are interleaved with a global transposition of the data in order to perform the FFTs along the remaining dimension. A well-known drawback of the slab decomposition strategy offered by FFTW is its limited parallel scalability, because at most $N$ MPI processes can be used for $N^3$ data.
We compensate for this by utilizing the hybrid MPI/OpenMP capability of FFTW (or functionally equivalent libraries such as Intel MKL), which allows to push the limits of scalability by at least an order of magnitude, corresponding to the number of cores of a modern multicore CPU or NUMA domain, respectively. All other relevant operations in the field solver can be straightforwardly parallelized with the help of OpenMP. Our newly developed parallel particle tracking algorithm has been implemented on top of this slab-type data decomposition using MPI and OpenMP, as shall be detailed below. Slab decompositions are beneficial for particle tracking since MPI communication overhead is minimized compared to, e.g., two-dimensional decompositions.

\subsection{Fluid solver}
\label{sec:fluid_solver}

The fluid solver consists of operations with field data, which \codename{} distributes among a total of $\numberMPIProcesses$ MPI processes with a standard slab decomposition, see Fig.~\ref{fig:slab_decomposition}.
Thus the logical field layouts consist of $(\numberGridNodes/\numberMPIProcesses) \times \numberGridNodes \times \numberGridNodes$ points for the real-space representation, and $(\numberGridNodes/\numberMPIProcesses) \times \numberGridNodes \times (\numberGridNodes/2+1)$ points for the Fourier-space representation. This allows the use of FFTW \cite{FFTW2005} to perform costly FFT operations, as outlined above.
We use the convention that fields are distributed along the real-space $x_3$ direction, and along the $k_2$ direction in the Fourier space representation (directions 2 and 3 are transposed between the two representations). Consequently, a problem on an $\numberGridNodes^3$ grid can be parallelized on a maximum of $\numberGridNodes$ computational nodes using one MPI process per node and, possibly, OpenMP threads inside the nodes, see Fig.~\ref{fig:slab_decomposition}.

In the interest of simplifying code development, \codename{} uses functional programming  for the costly traversal operation. Functional programming techniques allow to encapsulate field data in objects, while providing methods for traversing the data and computing specified arithmetic expressions --- i.e.~the class becomes a building block. While C++ allows for overloading arithmetic operators as a mechanism for generalizing them to arrays, our approach permits to combine several operations in a single data traversal, and it applies directly to operations between arrays of different shapes. In particular operations such as the combination of taking the curl and the Laplacian of a field (see \eqref{eq:NSVE}) are in practice implemented as a single field traversal operation.

\subsection{Particle tracking}
\label{sec:particletracersystem}

\begin{figure}
\includegraphics[width=\columnwidth]{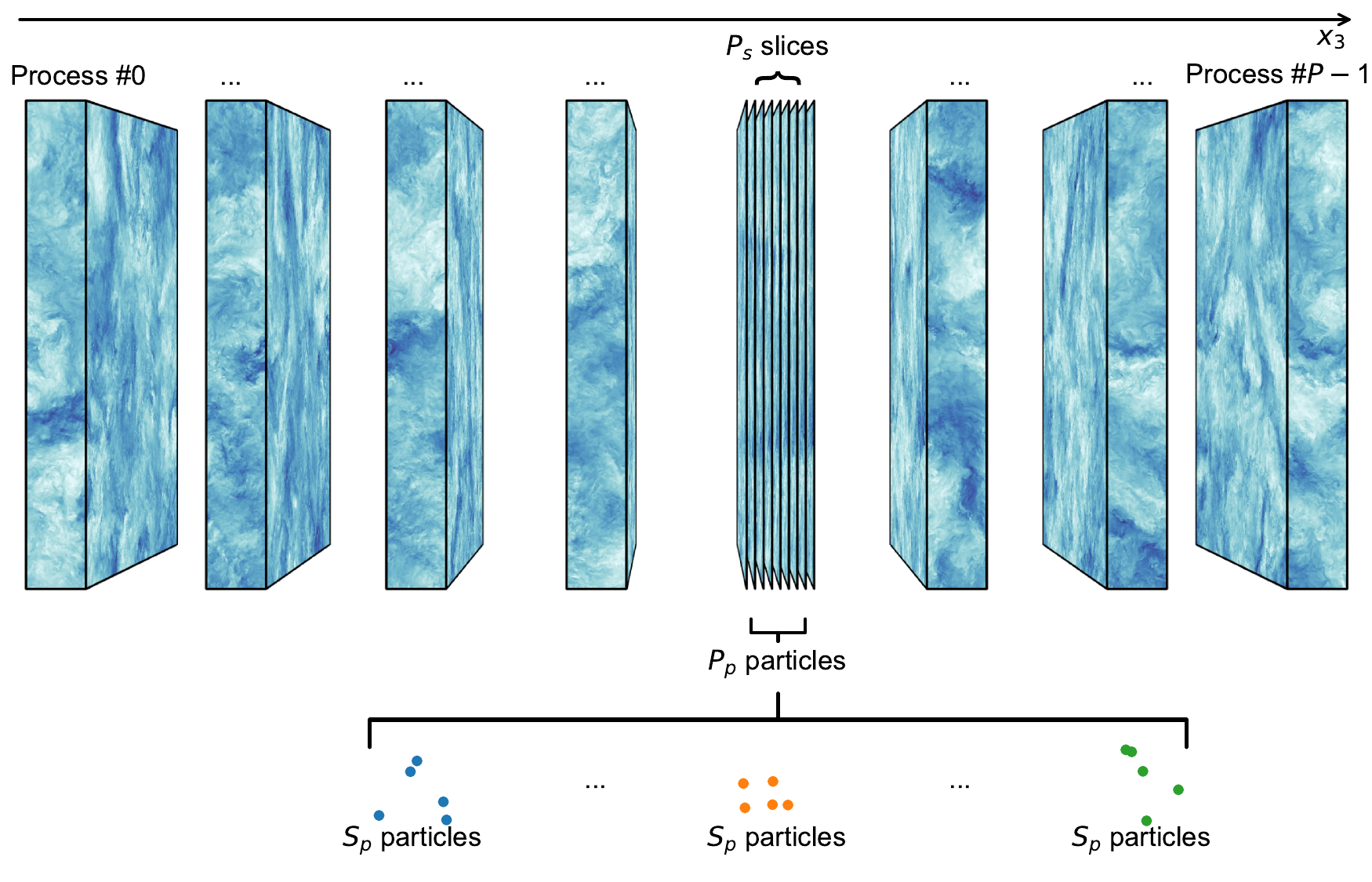}
\caption{
Distribution of real-space data between MPI processes in \codename{}.
Fields are split into \emph{slabs} and distributed between $\numberMPIProcesses$ MPI processes along the $x_3$ direction.
The $\numberParticles$ particles are also distributed, with each MPI process storing $\numberParticlesPerProcess$ particles on average.
Within each MPI process the particle data is sorted according to its $x_3$ location.
This leads to a direct association between each of the $\numberSlicesPerProcess$ field slices to contiguous regions of the particle data arrays --- in turn simplifying the interpolation procedure (see text for details).
On average, $\numberParticlesPerSlice$ particles are held within each such contiguous region.}
\label{fig:slab_decomposition}
\end{figure}

We now turn to a major feature of \codename{}: the efficient tracking of particles.
The novelty of our approach warrants a more in-depth presentation of the data structure and the parallel algorithms, for which we introduce the following notations (see also Fig.~\ref{fig:slab_decomposition}):

    \begin{itemize}
    \item $\numberMPIProcesses$ : the number of MPI processes (should be a
        divisor of the field grid size $\numberGridNodes$);
    \item $\numberSlicesPerProcess = \numberGridNodes/\numberMPIProcesses$ :
        the number of field slices in each slab;
    \item $\numberParticles$ : the number of particles in the system;
    \item $\numberParticlesPerProcess$: the number of particles contained in a given slab (i.e.
        hosted by the corresponding process) --- on average equal to $\numberParticles /
        \numberMPIProcesses$;
    \item $\numberParticlesPerSlice$: the number of particles per slice, i.e.~number
        of particles found between two slices --- on average equal to $\numberParticles / \numberGridNodes$;
    \item $\numberInterpolationKernelNodes$ : the width of the interpolation kernel, i.e.~the number
        of slices needed to perform the interpolation.
    \end{itemize}

The triple sum \eqref{eq:triple_sum} is effectively split into $\numberInterpolationKernelNodes$ double
sums over the $x_1$ and $x_2$ directions, the results of which then need to be distributed/gathered among the MPI processes such that the sum along the $x_3$ direction can be finalized.
Independently of $\numberMPIProcesses$ and $\numberGridNodes$, there will be $\numberParticles$ sums of $\numberInterpolationKernelNodes^3$ terms that have to be performed. However, the amount of information to exchange depends on the DNS parameters $\numberGridNodes$, $\numberParticles$, and $\numberInterpolationKernelNodes$, and on the job parameter $\numberMPIProcesses$.

Whenever more than one MPI process is used, i.e.~$\numberMPIProcesses > 1$, we distinguish between two cases:
    \begin{enumerate}
        \item $\numberInterpolationKernelNodes \leq \numberSlicesPerProcess$, i.e.~each MPI domain extends over at least as many slices as required for the interpolation kernel. In this case particles are shared between at most two MPI processes, therefore each process needs to exchange information with two other processes. In this case, the average number of shared particles is $\numberParticlesPerSlice (\numberInterpolationKernelNodes-1)$.
        \item $\numberInterpolationKernelNodes > \numberSlicesPerProcess$, i.e.~the interpolation kernel always extends outside of the local MPI domain. The average number of shared particles is $\numberParticlesPerSlice \numberSlicesPerProcess$. Each given particle is shared among a maximum of $\ceil*{\numberInterpolationKernelNodes/\numberSlicesPerProcess}$ processes, therefore each process must in principle communicate with $2\ceil*{\numberInterpolationKernelNodes/(2\numberSlicesPerProcess)}$ other processes.
    \end{enumerate}
The second scenario is the more relevant one for scaling studies.
Our expectation is that the communication costs will outweigh the computation costs, therefore the interpolation step should scale like $\numberParticles \numberInterpolationKernelNodes/\numberSlicesPerProcess = \numberParticles \numberInterpolationKernelNodes \numberMPIProcesses / \numberGridNodes$. In the worst case scenario, when the individual double sums have a significant cost as well, we expect scaling like $\numberParticles \numberInterpolationKernelNodes^3 \numberMPIProcesses / \numberGridNodes$.

\subsubsection{Particle data structure}
\label{sec:particle-storage}

The field grid is partitioned in one dimension over the processes, as described
in Section~\ref{sec:fluid_solver}, such that each process owns a field slab.
For each process, we use two arrays to store the data for particles included
inside the corresponding slab.
The first array contains state information, including the particle locations
--- required to perform the interpolation of the field.
We denote this first array \emph{state}.
The second array, called \emph{rhs}, contains the value of the right-hand side
of \eqref{eq:tracers}, as computed at the most recent few
iterations (as required for the Adams-Bashforth integration); updating this
second array requires interpolation.
The two arrays use an array of structures pattern, in the sense that data
associated to one particle is contiguous in memory.
While this may lead to performance penalties, as pointed out in \cite{homann2018cpc}, there are significant benefits for our MPI parallel approach, as explained below.
We summarize in the following the main operations that are applied to the arrays.

\paragraph{Ordering the particles locally}

When $\numberGridNodes > \numberMPIProcesses$, processes are in charge of more
than one field slice, and the particles in the slab are distributed across different slices.
In this case, we store the particles that belong to the same slice contiguously
in the arrays, one slice after the other in increasing $x_3$-axis order.
This can be achieved by partitioning the arrays into $\numberSlicesPerProcess$
different groups (indexed by $s$ in the following) and can be implemented as an incomplete Quicksort with a
complexity of $O(\numberParticlesPerProcess \, \log \, \numberSlicesPerProcess)$
on average.
The different groups will in general contain different numbers of particles, which implies that the associated data is stored at arbitrary locations within the contiguous \emph{state} and \emph{rhs} arrays.
We therefore build an array \emph{offset} of size $\numberSlicesPerProcess+1$, where \emph{offset[s]} returns the starting index of the first particle for the group \emph{s} and the number of particles in this same group \emph{s} can be computed as \emph{offset[s+1]}-\emph{offset[s]}.
For simplicity, we explicitly store \emph{offset[$\numberSlicesPerProcess$]}$=\numberParticlesPerProcess$ as the $\numberSlicesPerProcess+1$-th element of \emph{offset}.
This allows direct access to the contiguous data regions corresponding to each field slice, access which is in turn relevant for MPI exchanges (see below).

\paragraph{Exchanging the particles for computation}
With our data structures, we are able to send the state information of all the particles located in a single group with only one communication, which reduces communication overhead.
Moreover, sending the particles from several contiguous groups can also be done in a single operation because the groups are stored sequentially inside the arrays.

\paragraph{Particles displacement/update}
The positions of the particles are updated at the end of each iteration, and so
the arrays must be rearranged accordingly.
The changes in the $x_3$ direction might move some particles to a different slice
and even to a slice owned by a different process.
Therefore, we first partition the first and last groups (the groups of the first
and last slices of the process's slab) to move the particles that are now outside
of the process's grid interval at the extremities of the arrays.
We only act on the particles located at the lower and higher groups because we
assume that the particles cannot move a distance greater than
$2\pi/\numberGridNodes$. For regular tracers \eqref{eq:tracers} this is in
fact required by the CFL stability condition of the fluid solver.
This partitioning is done with a complexity $O(\numberParticlesPerProcess /
\numberSlicesPerProcess)$.
Then, every process exchanges those particles with its direct neighbors, ensuring that the particles are correctly distributed.
Finally, each process sorts its particles to take into account the changes in
the positions and the newly received particles as described previously.

\subsubsection{Parallelization}

\begin{figure*}[ht]
  \centering
  \includegraphics[width=\textwidth, keepaspectratio, page=1]{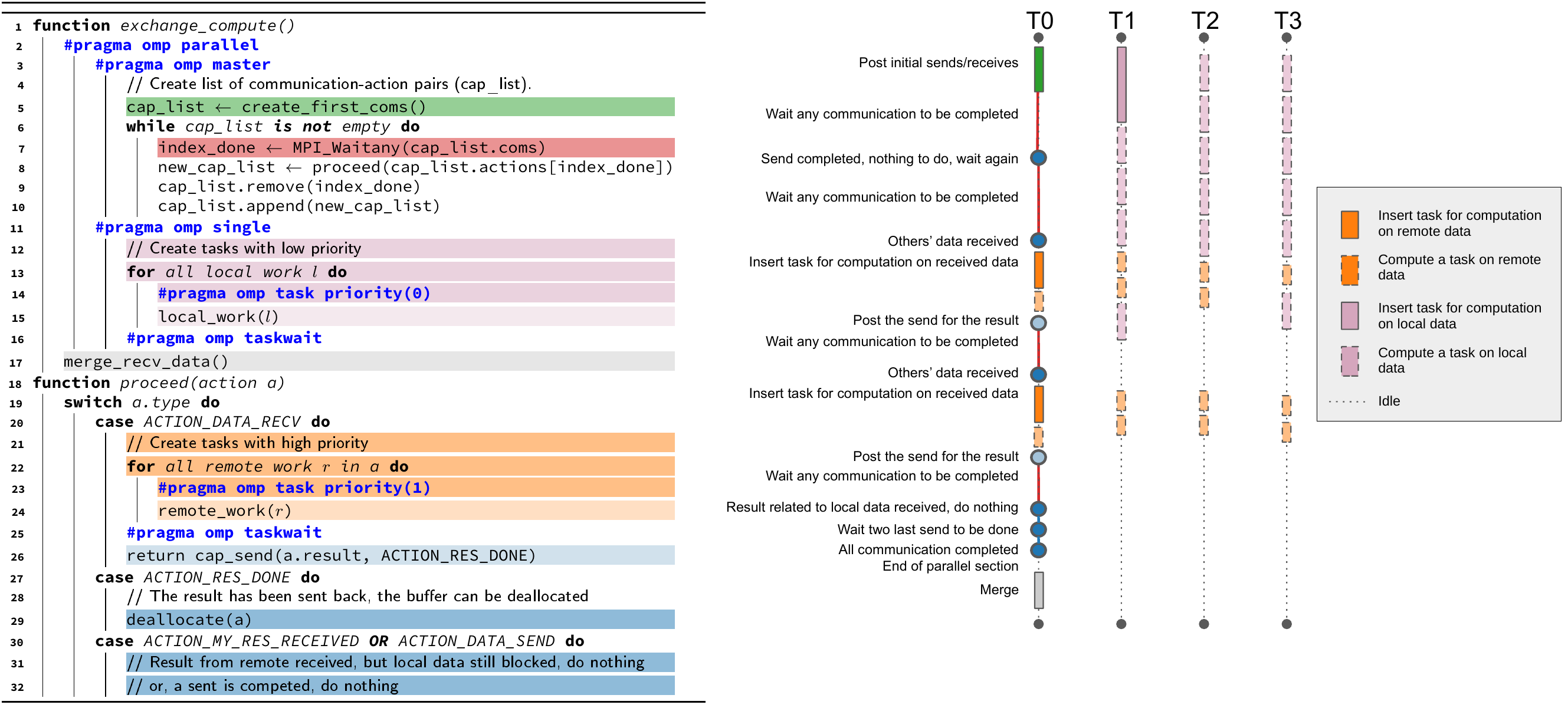}
  \caption{Overview of parallel interpolation algorithm (left) and reference execution timeline for one process with four threads (right).
  Colored lines of code correspond to individual timeline elements.
  The threads compute local operations by default but switch to remote operations when the master thread creates the corresponding new higher-priority tasks.
  With the use of priorities, the execution pattern allows for quicker communication of results to different processes.
  See text for details.
  }
  \label{fig:execution}
\end{figure*}

The interpolation of the field at the particle locations concentrates most of
the workload of the numerical particle tracking.
For each particle, the interpolation uses the $\numberInterpolationKernelNodes^3$ surrounding field nodes.
However, because we do not mirror the particle or the field information on multiple processes, we must actively exchange either field or particle information to perform a complete interpolation.
Assuming that the number of particles in the simulation is much smaller than the
number of field nodes, i.e.~the relation $\numberParticlesPerProcess <
\numberInterpolationKernelNodes \numberGridNodes^2$ holds, less data needs to be transferred on average when particle locations are exchanged rather than field values at required grid nodes.
Consequently, in our implementation we exchange the particle information only.

A straightforward implementation in which the communication and computation are dissociated consists in the following operations:
\begin{enumerate}[(a)]
\item each process computes the interpolation of its particles on its field;
\item all the processes exchange particle positions with their neighbors (each process sends and receives arrays of positions);
\item each process computes the interpolation using its field on the particle positions it received from other processes in (b);
\item all the processes exchange the results of the interpolations from (c) with the corresponding neighbors;
\item each process merges the results it received in (d) and the results from its own computation from (a).
\end{enumerate}

In our implementation, we interleave these five operations to overlap communication with computation. As we detail in the following, the master thread of each MPI process creates computation work packages, then performs communications while the other threads are already busy with the work packages.
This is achieved with the use of non-blocking MPI communications and OpenMP tasks, as illustrated in Fig.~\ref{fig:execution}.
Broadly speaking, the execution is guided by a pairing of communications with relevant actions that need to be performed on the associated data (either sent or received), with all such pairs put together in a list \nameOfCAPList{}  (\texttt{cap\_list} in Fig.~\ref{fig:execution}).
Actions of "compute" type are turned into tasks distributed to all available threads, whereas actions of "send/receive" type are handled separately by the master thread.
A more detailed overview follows.
The figure shows a simplified version of the parallel execution pattern (the function \texttt{exchange\_compute}), along with a possible execution timeline.
For simplicity we also define the function \texttt{proceed}, which encompasses much of the communication and coordination activity of the master thread.
Here we focus on describing the algorithm, only referring to the execution timeline implicitly through the color correspondence.
In a first stage, the master thread splits the local interpolation from (a) into tasks and submits them immediately but with a low priority.
Then, it posts all the sends/receives related to (b) and all the receives related to (d), and creates the list \nameOfCAPList{}.
In the core part of the algorithm, the master thread performs a wait-any on the list of MPI requests (line 7 in the pseudocode).
This MPI function is blocking and returns as soon as one of the communications in the list is completed.
Hence, when a communication is completed, the master thread acts accordingly to the type of action \nameOfAction{} it represents (\texttt{a.type} in Fig.~\ref{fig:execution}).
If \nameOfAction{} is the completion of a send of local particle positions, from (b), there is nothing to do and the master thread directly goes back to the wait-any.
In this case, it means that a send is completed and that there is nothing new to do locally.
If \nameOfAction{} is the completion of a receive of remote particle positions, from (b), then the master thread creates tasks to perform the interpolation of these positions, from (c), and submits them with high priority (lines 23 and 24 in the pseudocode).
Setting a high priority ensures that all the threads will work on these tasks even if the tasks inserted earlier to interpolate the local positions, from (a), are not completed.
When these tasks are completed, the master thread posts a non-blocking send to communicate the results to the process that owns the particles and adds the corresponding MPI request to \nameOfCAPList{}.
Then, the master thread goes back to the wait-any on \nameOfCAPList{}.
If \nameOfAction{} is the completion of a send of interpolation on received positions, as just described, the master thread has nothing to do and goes back to the wait-any on \nameOfCAPList{}.
In fact, this event simply means that the results were correctly sent.
If \nameOfAction{} is the completion of a receive, from (d), of interpolation performed by another process, done in (c), the master thread keeps the buffer for merging at the end, and it goes back to the wait-any.
When \nameOfCAPList{} is empty, it means that all communications (b,d) but also computations on other positions (c) are done.
If some local work still remains from (a), the master thread can join it and compute some tasks.
Finally, when all computation and communication are over, the threads can merge the interpolation results, operation (e) (line 17 in the pseudocode).

The described strategy is a parallel programming pattern that could be applied in many other contexts when there are local and remote works to perform and where remote work means first to exchange information and second to apply computation on it.

\subsection{In-Order Parallel Particle I/O}
\label{sec:inordersaving}

\begin{figure}[h!]
   \centering
   \includegraphics[width=\columnwidth, height=.4\textheight, keepaspectratio, page=1]{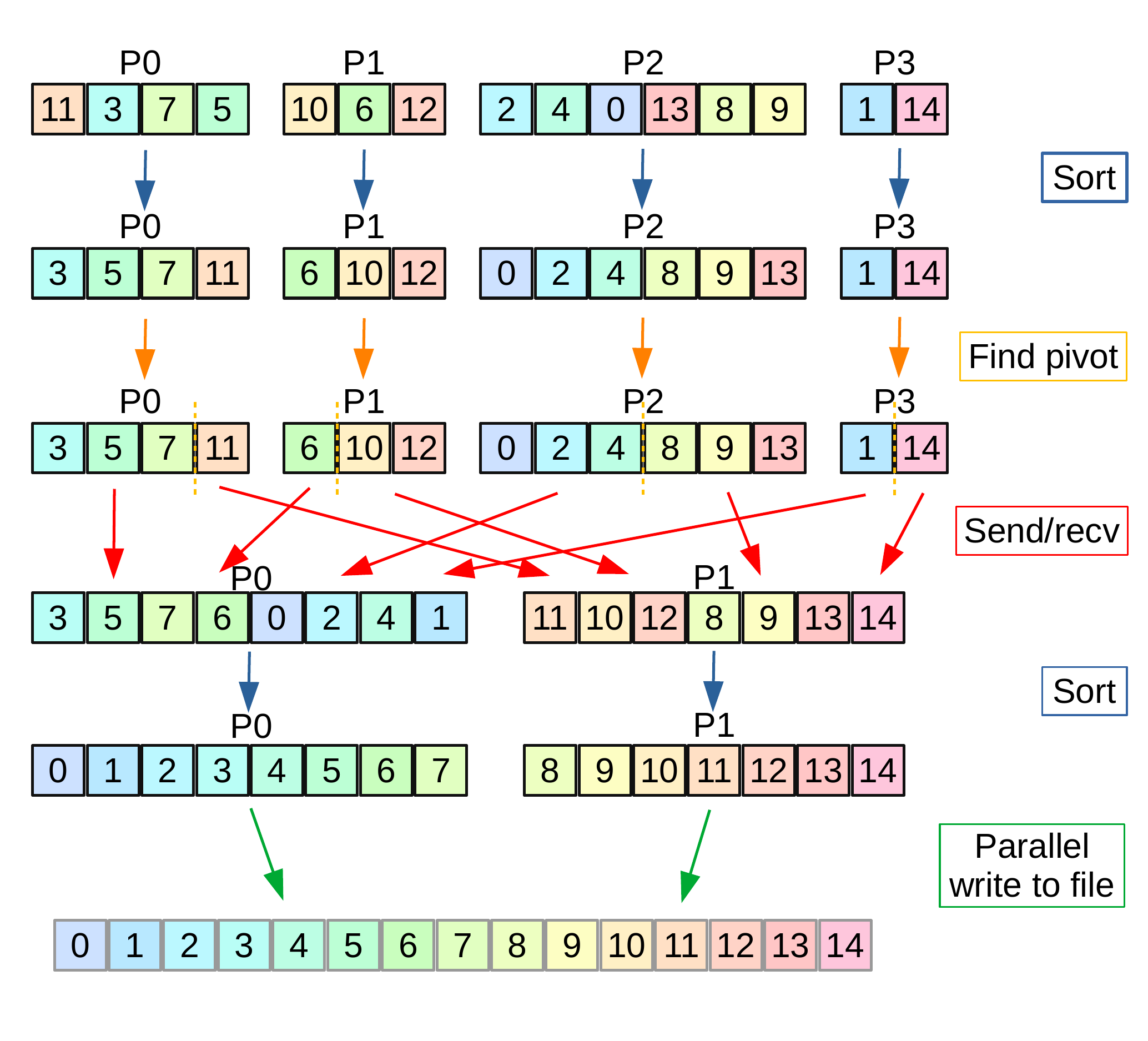}
   \caption{The different stages to perform a parallel saving of the particles in order.
            Here, we consider that the particle data (illustrated by the global particle index) is distributed among 4 processes, but that only 2 of them are used in the write operation.
            }
   \label{fig:io}
\end{figure}

Saving the states of the particles on disk is a crucial operation to support checkpoint/restart and for post-processing.
We focus on output because \codename{} typically performs many more output than input operations (the latter only happen during initialization).
The order in which the particles are saved is important because it influences the writing pattern and the data accesses during later post-processing of the files.
As the particles move across the processes during the simulation, a naive output of the particles as they are distributed will lead to inconsistency from one output to the other.
Such a structure would require reordering the particles during the post-processing or would result in complex file accesses.
That is why we save the particles in order,  i.e.~in the original order given as input to the application.

The algorithm that we use to perform the write operation is shown in Fig.~\ref{fig:io}.
There are four main steps to the procedure: \emph{pre-sort} ("Sort" and "Find pivot" in the figure), followed by \emph{exchange} ("Send/Recv" in Fig.~\ref{fig:io}), with a final \emph{post-sort} ("Sort") before the actual HDF5 \emph{write} ("Parallel write to file" in the figure).

\begin{figure*}[t]
    \includegraphics[width=\textwidth]{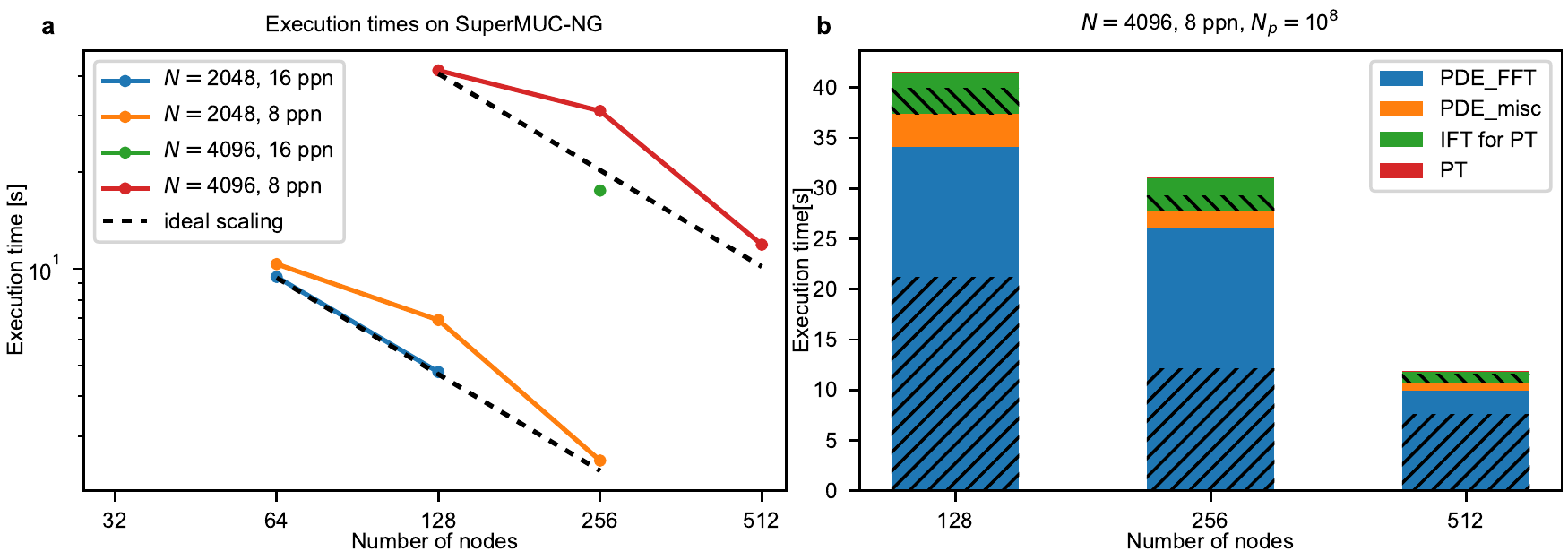}
    \caption{Computational performance of \codename{}. Strong scaling behavior of
    the total runtime (a) for grids with $\numberGridNodes=2048$ and
    $\numberGridNodes=4096$, respectively, and using 8 or 16 MPI processes per
    node (ppn) on up to 512 fully populated nodes (24576 cores) of SuperMUC-NG
    ($\numberInterpolationKernelNodes=8$, $\numberParticles=10^8$).
    For $\numberGridNodes=4096$ and 8 ppn panel (b) shows a
    breakdown of the total runtime into the main functional elements,
    namely solving the system of Navier-Stokes partial differential
    equations ("PDE\_misc" and "PDE\_FFT") and particle tracking ("IFT for PT" and "PT").
    The fluid solver cost is largely dominated by the fast Fourier
    transforms ("PDE\_FFT"). The cost of particle tracking for $10^8$
    particles (with $\numberInterpolationKernelNodes=8$) is determined
    by the additional inverse Fourier transform ("IFT for PT"), whereas the
    runtime for our novel particle tracking algorithm ("PT") is still
    negligible for $10^8$ particles. Hatched regions represent the fraction of MPI
    communication times.}
    \label{fig:overall_performance}
\end{figure*}

Each process first sorts its local particles using the global indices with $O(\numberParticlesPerProcess \, \log \, \numberParticlesPerProcess)$ complexity.
This sort can be done in parallel using multiple threads.
Then, each process counts the number of particles it has to send to each of the processes that are involved in the file writing.
These numbers are exchanged between the processes allowing each process to allocate the reception buffer.
If we consider that $\numberParticleOutputProcesses$ processes are involved in the output operation, each of them should receive
$\numberParticles/\numberParticleOutputProcesses$ particles in total, and a process of rank $r$ should receive the particles from index $r \times
\numberParticles/\numberParticleOutputProcesses$ to $(r+1) \times \numberParticles / \numberParticleOutputProcesses - 1$.

In the \emph{exchange} step, the particles can be sent either with multiple non-blocking send/receive operations or with a single all-to-all operation, with the total number of communications bounded by $\numberMPIProcesses \times \numberParticleOutputProcesses$.
Finally, the received particles are sorted with a complexity of
$O(\numberParticles/\numberParticleOutputProcesses \, \log \,
\numberParticles/\numberParticleOutputProcesses)$, and written in order into the output file.

The number $\numberParticleOutputProcesses$ of processes involved in the writing should be carefully chosen because
as $\numberParticleOutputProcesses$ increases, the amount of data output per process decreases and might become so small that the write operation becomes inefficient. At the same time, the preceding exchange stage becomes more and more similar to a complete all-to-all communication with $N_p^2$ relatively small messages.
On the other hand, as $\numberParticleOutputProcesses$ decreases, the size of the messages exchanged will increase, and the write operation can eventually become too expensive for only a few processes, which could also run out of memory. This is why we heuristically fix $\numberParticleOutputProcesses$ using three parameters: the minimum amount of data a process should write, the maximum number of processes involved in the write operation, and a chunk size.
As $\numberParticles$ increases, $\numberParticleOutputProcesses$ increases up to the given maximum.
If $\numberParticles$ is large enough, the code simply ensures that $\numberParticleOutputProcesses-1$ processes output the same amount of data (being a multiple of the chunk size), and the last process writes the remaining data.
In our implementation, the parameters are chosen empirically (based on our experience with several HPC clusters running the IBM GPFS/SpectrumScale parallel file system), and they can be tuned for specific hardware configurations if necessary.

We use a similar procedure for reading the particle state: $\numberParticleOutputProcesses$ processes read the data, they sort it according to spatial location, then they redistribute it to all MPI processes accordingly.

\section{Computational performance}

\begin{figure*}[t]
    \includegraphics[width=\textwidth]{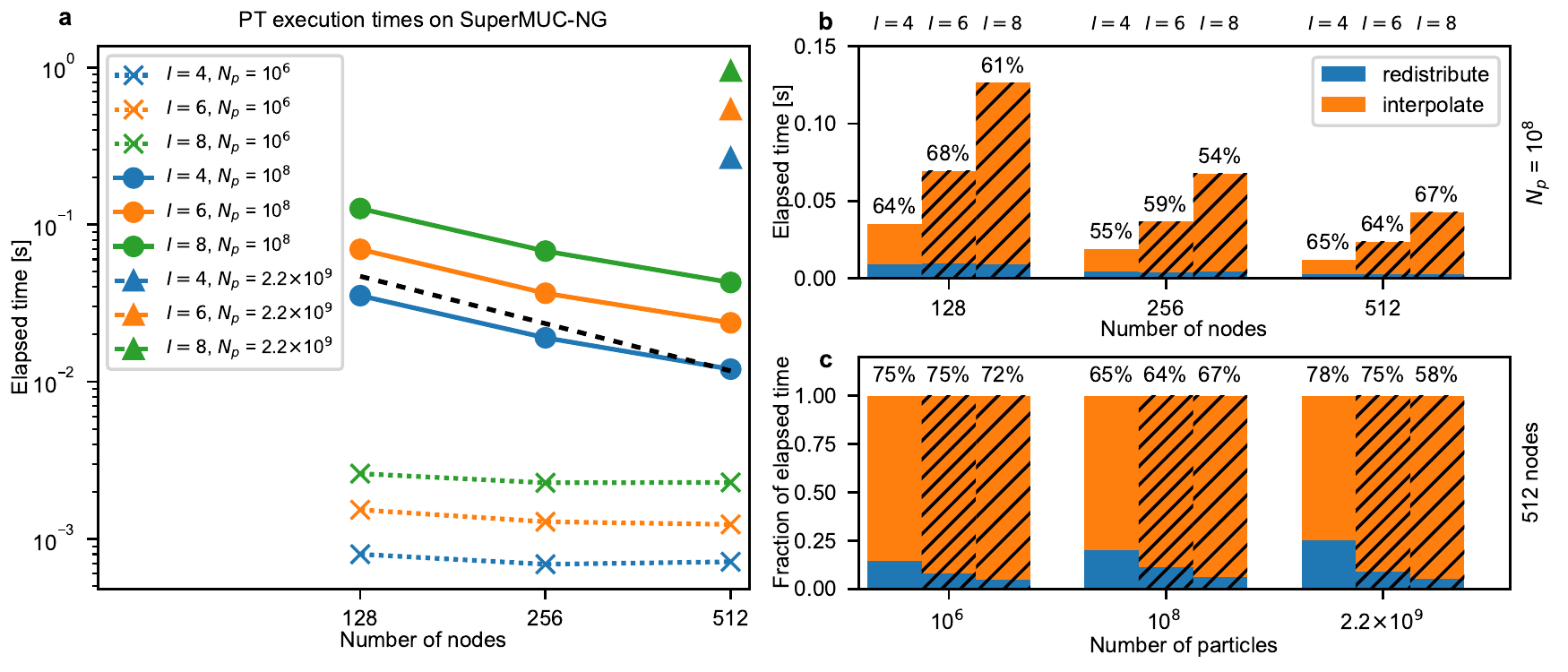}
    \caption{Computational performance of the particle tracking code using 8 MPI processes per node and a DNS of size $\numberGridNodes=4096$, for different sizes of the interpolation kernel $\numberInterpolationKernelNodes$. Panel (a): strong scaling for different numbers of particles $\numberParticles$ and sizes of the interpolation kernel (memory requirements limit the $\numberParticles = 2.2 \times 10^9$ case to 512 nodes). The dashed line corresponds to ideal strong scaling. Panel (b): contributions of interpolation and redistribution operations to the total execution time, for a fixed number of particles, $\numberParticles = 10^8$, and for different sizes of the interpolation kernel (the corresponding vertical bars are distinguished by hatch style, see labels on top) as a function of the number of compute nodes. Panel (c): relative contributions of interpolation and redistribution as a function of $\numberParticles$. Percentages represent the fraction of time spent in MPI calls.}
    \label{fig:particle_performance}
\end{figure*}

\label{sec:performance}
\subsection{Hardware and software environment}

To evaluate the computational performance of our approach, we perform benchmark simulations on the HPC cluster SuperMUC-NG from the Leibniz Supercomputing Centre (LRZ):
we use up to 512 nodes containing two Intel Xeon Platinum 8174 (Skylake) processors with 24 cores each and a base clock frequency of 3.1 GHz, providing 96 GB of main memory.
The network interconnect is an Intel OmniPath (100 Gbit/s) with a pruned fat-tree topology that enables non-blocking communications within islands of up to 788 nodes. 
We use the Intel compiler 19.0, Intel MPI 2019.4, HDF5 1.8.21 and FFTW 3.3.8.
For our benchmarks, we always fully populate the nodes, i.e.~the combination of MPI processes per node (ppn) and OpenMP threads per MPI process is chosen such that their product equals 48, and that the threads spawned by an MPI rank are confined within the NUMA domain defined by a single processor.  

\subsection{Overall performance}

\begin{table}[bh]
\begin{small}
\begin{center}
\begin{tabular}{c|c|c|c|c}
    &
    \multicolumn{2}{c}{N = 2048} &
    \multicolumn{2}{c}{N = 4096}\\
    \hline
    \diagbox[width=10em]{Number\\of nodes}{ppn} &
    8 & 16 &
    8 & 16 \\
    \hline
    64 & 10.3 & 9.43 & --- & --- \\
    128 & 6.92 & 4.76 & 41.6 & --- \\
    256 & 2.52 & --- & 31.1 & 17.5 \\
    512 & --- & --- & 11.9 & ---
\end{tabular}
\end{center}
\end{small}
\caption{Execution times for TurTLE on SuperMUC-NG (in seconds), various configurations (values correspond to Fig.~\ref{fig:overall_performance}a).}
\label{tab:overall_performance_a}
\end{table}

Figure~\ref{fig:overall_performance} provides an overview of the
overall parallel scaling behavior of the code for a few typical
large-scale setups (panel a as well as Table~\ref{tab:overall_performance_a}) together with a breakdown into the
main algorithmic steps (panel b). We use the execution time
for a single time step (averaged over a few steps) as the primary
performance metric and all data and computations are handled in double precision.
The left panel shows, for two
different setups ($\numberGridNodes = 2048, 4096$),
that the code exhibits excellent strong-scaling efficiency (the dashed
line represents ideal scaling) from the
minimum number of nodes required to fit the code into memory
up to the upper limit which is given by the maximum number of MPI processes
that can be utilized with our one-dimensional domain decomposition.  
The total runtime of TurTLE, and hence its major parallel scaling properties are largely determined
by the three-dimensional fast Fourier transforms (using FFTW) for solving the
system of Navier Stokes partial differential equations, as shown in Fig.~\ref{fig:overall_performance}b (labeled "PDE\_FFT", entire blue area) for the example of the large setup ($\numberGridNodes = 4096$) with 8 processes per node (see also Table~\ref{tab:overall_performance_b}). With increasing node count, the runtime of the FFTs in turn gets increasingly dominated by an all-to-all type of MPI communication pattern which is arising from the FFTW-internal global transpositions (blue-hatched area) of the slab-decomposed data.
\begin{table}[bh]
\begin{small}
\begin{center}
\begin{tabular}{c|S[table-format=1.0]r|S[table-format=1.0]r|S[table-format=1.1]r p{4cm}}
    \diagbox[width=9em]{Algorithm unit}{Number\\of nodes} &
    \multicolumn{2}{c}{128} &
    \multicolumn{2}{c}{256} &
    \multicolumn{2}{c}{512} \\
    \hline
    PDE\_FFT   & 34    &(62\%)& 26    &(46\%)& 9.9   &(76\%)\\
    PDE\_misc  &  3.3  & (-\%)&  1.7  & (-\%)& 0.74  & (-\%)\\
    IFT for PT &  4.1  &(62\%)&  3.3  &(46\%)& 1.2   &(76\%)\\
    PT         &  0.13 &(61\%)&  0.07 &(55\%)& 0.043 &(67\%)
\end{tabular}
\end{center}
\end{small}
\caption{Breakdown of the total TurTLE runtime (in seconds) into main functional elements (values correspond to Fig.~\ref{fig:overall_performance}b).}
\label{tab:overall_performance_b}
\end{table}

While a good OpenMP efficiency can be noted for the case of 64 nodes
by comparing the blue and the orange curve in Fig.~\ref{fig:overall_performance}a, i.e.~the same problem
computed with a different combination of MPI processes per node (8, 16)
and a corresponding number of OpenMP threads per process (6, 3), we generally observe
FFTW exhibiting limited OpenMP scaling beyond a number of 6 to 8 threads per 
MPI process for the grid resolutions ($N$) considered here. Moreover, even at moderate
thread counts, FFTW, due to internal load-imbalances, fails to efficiently handle the process-local transforms for
certain dimensions of the local data slabs: In the case of 256 nodes, FFTW apparently
cannot efficiently use more than 3 OpenMP threads for parallelizing
over the local slabs of dimension $2 \times 4096 \times 2049$ (Fourier representation), whereas good scaling
up to the desired maximum of 6 threads is observed for a dimension of
$8 \times 4096 \times 2049$ (128 nodes) and also $1 \times 4096 \times 2049$ (512 nodes).
The same argument applies for the smaller setup ($\numberGridNodes = 2048$)
on 128 nodes and is the root cause for the kink in the strong scaling curves of panel a. 
We plan for TurTLE to support FFTs also from the Intel Math Kernel Library (MKL) which are expected to deliver improved threading efficiency.
For practical applications, this is less of a concern, since a user can perform a few exploratory
benchmarks for a given setup of the DNS on the particular
computational platform, and available node counts, in order to find an
optimal combination of MPI processes and OpenMP threads. Since the
runtime per time step is constant for our implementation of the
Navier-Stokes solver, a few time steps are sufficient for tuning a
long-running DNS.

Thanks to our efficient and highly scalable implementation of the
particle tracking, its contribution to the total runtime is barely
noticeable in the figure ("PT", red colour in
Fig.~\ref{fig:overall_performance}b). This holds
even for significantly larger numbers of particles than the value of $N_p=10^8$ which was used here (see below for
an in-depth analysis). 
The only noticeable additional cost for particle tracking, amounting to
roughly 10\% of the total runtime, comes from an additional inverse FFT
("IFT for PT", green colour) which is required to compute the
advecting velocity field, which is independent of $N_p$ and scales very well.

Finally, Fig.~\ref{fig:overall_performance}a also suggests good weak scaling behavior of \codename{}: 
When increasing the resolution from $\numberGridNodes = 2048$ to
$\numberGridNodes = 4096$ and at the same time increasing the number
of nodes from 64 to 512, the runtime increases from 10.35s to 11.45s, which is consistent with
a $O(\numberGridNodes^3 \log \numberGridNodes)$ scaling of the
runtime, given the dominance of the FFTs. 

\subsection{Particle tracking performance}

Fig.~\ref{fig:particle_performance} provides an overview and some
details of the performance of our particle tracking algorithm,
extending the assessment of the previous subsection to particle
numbers beyond the current state of the art \cite{buaria2017cpc}.
We use the same setup of a DNS with $N=4096$ and 8 MPI processes per node on SuperMUC-NG, as presented in the previous subsection. 

Fig.~\ref{fig:particle_performance}a summarizes the strong-scaling
behavior on 128, 256 or 512 nodes of SuperMUC-NG for different
numbers of particles ($10^6$, $10^8$ and $2.2 \times 10^{9}$) and for different
sizes of the interpolation kernel $\numberInterpolationKernelNodes$ (4, 6, 8).
Most importantly, the absolute run times are small compared to the fluid
solver: Even when using the most accurate interpolation kernel,
a number of $2.2 \times 10^{9}$ particles can be handled within less than a second (per time step), i.e.~less than 10\% to the total computational cost of \codename{} on 512 nodes per time step (cf. Fig.~\ref{fig:overall_performance}). 

The case of $\numberParticles = 10^6$ is shown only for reference
here. This number of particles is too small to expect good
scalability in the regime of 128 compute nodes and more. Still, the
absolute runtimes are minuscule compared to a DNS of typical size.
For $\numberParticles = 10^8$ we observe good but not perfect strong
scaling, in particular for the largest interpolation kernel ($\numberInterpolationKernelNodes=8$),
suggesting that we observe the $\numberParticles \numberInterpolationKernelNodes \numberMPIProcesses / \numberGridNodes$ regime, as discussed previously.
It is worth mentioning that we observe a sub-linear scaling of the total
runtime with the total number of particles (Fig.~\ref{fig:particle_performance}a).

Fig.~\ref{fig:particle_performance}b shows a
breakdown of the total runtime of the particle tracking algorithm into
its main parts, interpolation (operations detailed in
Fig.~\ref{fig:execution}, shown in orange) and redistribution (local sorting of
particles together with particle exchange, blue), together with
the percentage of time spent in MPI calls. The latter takes between half and
two thirds of the total runtime for $\numberParticles =
10^8$ particles (cf. upper panel b) and reaches almost 80\% for
$\numberParticles = 2.2 \times 10^9$ particles on 512 nodes
(lower panel c).
Overall, the interpolation cost always dominates over redistribution,
and increases with the size of the interpolation kernel roughly
as $\numberInterpolationKernelNodes^2$, i.e.~the interpolation cost is proportional to the number of MPI messages required by the algorithm (as detailed above).

\subsection{Particle output}

\begin{figure}[tbh]
    \centering
    \includegraphics[width=\columnwidth]{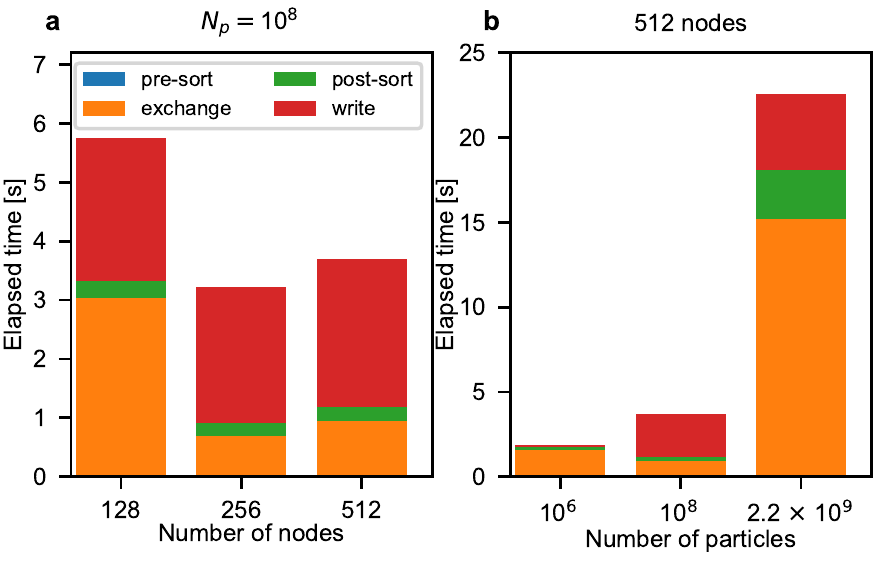}
    \caption{Performance of particle output, as distributed between the four different operations (see also Fig.~\ref{fig:io}): \textit{pre-sort} (particles are sorted by each process), MPI \textit{exchange} (particle data is transferred to processes actually participating in I/O), \textit{post-sort} (particles are sorted on each I/O process), and \textit{write} (HDF5 write call).
    Panel (a): elapsed times as a function of the total number of nodes, for a fixed $\numberParticles = 10^8$ (see also Fig.~\ref{fig:particle_performance}b). 
    Panel (b): elapsed time as a function of the number of particles, in the case of 512 nodes (see also Fig.~\ref{fig:particle_performance}c).
    }
    \label{fig:particle_output}
\end{figure}

Figure~\ref{fig:particle_output} provides an overview of the computational costs of the main parts of the output algorithm,
namely sorting particles according to their initial order (\textit{pre-sort} and \textit{post-sort} stages, cf.~Sect.~\ref{sec:inordersaving}),
communicating the data between processes (\textit{exchange} stage), and writing data to disk using parallel HDF5 (\textit{write} stage).
Here, the same setup is used as in Fig.~\ref{fig:particle_performance} panels b and c, respectively, noting
that the output algorithm does not depend on the size of the interpolation kernel.
The figure shows that the total time is largely dominated by the \textit{write} and \textit{exchange} stages, with the sorting stages not being significant. Of the latter, the \textit{post-sort} operation is relatively more expensive than the \textit{pre-sort} stage,
because only a comparably small subset of $\numberParticleOutputProcesses < \numberMPIProcesses$ processes is used in the \textit{post-sort} stage
(in the present setup $\numberParticleOutputProcesses =1$ for $10^6$ particles, $\numberParticleOutputProcesses =72$ for $10^8$ particles,
and $\numberParticleOutputProcesses =126$ for $2.2\times 10^9$ particles were used).
This indicates that our strategy of dumping the particle data in order adds only a small overhead, which is mostly spent in the communication stage (unsorted output
could be done with a more simple communication pattern) but not for the actual (process-local) reordering of the particles.
For a given number of particles $\numberParticles$, the number of processes $\numberParticleOutputProcesses$ involved in the \textit{write} operation
is fixed, independent of the total number $\numberMPIProcesses$ of processes used for the simulation. Consequently, the time spent in the 
\textit{write} stage does not depend on the number of nodes (and hence $\numberMPIProcesses$), as shown in Fig.~\ref{fig:particle_output}a.
However, $\numberParticleOutputProcesses$ may increase with increasing $\numberParticles$ (and fixed $\numberMPIProcesses$).

Fig.~\ref{fig:particle_output}b shows that the cost of writing $10^6$ particles with a single process is negligible, whereas writing $10^8$ particles with 72 processes becomes significant, even though a similar number of particles per output process ($1.4 \times 10^6$ particles) is used. This reflects the influence of the (synchronization) overhead of the parallel HDF5 layer and the underlying parallel IO system.
On the other hand, it takes about the same amount of time for 126 processes to write $1.7\times 10^7$ particles each, compared with 72 processes writing $1.4\times 10^6$ particles each, which motivates our strategy of controlling the number of processes $\numberParticleOutputProcesses$ that are involved in the interaction with the IO system.
However, the choice of $\numberParticleOutputProcesses$ also influences the communication time spent in the \textit{exchange} stage.
When looking at the \textit{exchange} stage in Figure~\ref{fig:particle_output}a, we recall that 72 processes write the data for all three node counts. As $\numberMPIProcesses$ increases, the 72 processes receive less data per message but communicate with more processes.
From these results it appears that this is beneficial: reducing the size of the messages but increasing the number of processes that communicate reduces the overall duration of this operation (that we do not control explicitly since we rely on the \texttt{MPI\_Alltoallv} collective routine).
For a fixed number of processes and an increasing number of particles (see Figure~\ref{fig:particle_output}b), the total amount of data exchanged increases and the size of the messages varies. The number $\numberParticleOutputProcesses$ (i.e., 1, 72 and 126) is not increased proportionally with the number of particles $\numberParticles$ (i.e., $10^6$, $10^8$ and $2.2\times 10^9$), which means that the messages get larger and, more importantly, each process needs to send data to more output processes.
Therefore, increasing $\numberParticleOutputProcesses$ also increases the cost of the \textit{exchange} stage but allows to control the cost of \textit{write} stage.
Specifically, it takes about 4s to output $10^8$ particles (1s for \textit{exchange} and 3s for \textit{write}). It takes only 6 times longer, about 23s (15s for \textit{exchange}, 4s for \textit{write}, and 3s \textit{post-sort}) to output 22 times more, $2.2\times 10^9$, particles. 

Overall, our strategy of choosing the number of processes $\numberParticleOutputProcesses$ participating in the IO operations independent of the total number $\numberMPIProcesses$ of processes allows us to avoid performance-critical situations where too many processes would access the IO system, or too many processes would write small pieces of data. The coefficients used to set $\numberParticleOutputProcesses$ can be adapted to the specific properties (hardware and software stack) of an HPC system.

\section{Summary and conclusions}

In the context of numerical studies of turbulence, we have presented a novel particle tracking algorithm using an MPI/OpenMP hybrid programming paradigm.
The implementation is part of \codename{}, which uses a standard pseudo-spectral approach for the direct numerical simulation of turbulence in a 3D periodic domain.
\codename{} succeeds at tracking billions of particles with a negligible cost relative to solving the fluid equations.
MPI communications are overlapped with computation thanks to a parallel programming pattern that mixes OpenMP tasks and MPI non-blocking communications.
At the same time, the use of a contiguous and slice-ordered particle data storage allows to minimize the number of required MPI messages for any size of the interpolation kernel.
This way, our approach combines both numerical accuracy and computational performance to address open questions regarding particle-laden flows by performing highly resolved numerical simulations on large supercomputers\cite{lalescu2018njp, bentkamp2019ncomms, pujara2021jfm, bentkamp2022nc}.
Indeed, as the benchmarks presented in Section 4 have shown, \codename{} scales up to many thousands of CPU cores on modern high-performance computers.

We expect that due to our task-based parallelization and the asynchronous communication scheme the particle-tracking algorithm is also well suited for offloading to massively parallel accelerators (e.g.~GPUs) of a heterogeneous HPC node architecture.
Concerning the fluid solver, the pseudo-spectral approach taken by \codename{} appears in principle amenable to GPU acceleration, as demonstrated by similar fluid codes (e.g. \cite{mukherjee2018hipcw, ravikumar2019GPUAO, lopez2020softwarex, rosenberg2020atmosphere}).
Whether it can be implemented in a (performance\nobreakdash-) portable and scalable way on large GPU-accelerated HPC systems remains to be investigated.

\section{Acknowledgments}
The authors gratefully acknowledge the Gauss Centre for Supercomputing e.V. (www.gauss-centre.eu) for funding this project by providing computing time on the GCS Supercomputer SuperMUC-NG at Leibniz Supercomputing Centre (www.lrz.de).
Some computations were also performed at the Max Planck Computing and Data Facility. This work was supported by the Max Planck Society. This work is part of a project that has received funding from the European Research Council (ERC) under the European Union’s Horizon 2020 research and innovation programme (Grant agreement No. 101001081).

\end{document}